\documentclass[a4paper]{jpconf}

\usepackage{graphicx}
\usepackage{dcolumn}
\usepackage{bm}
\usepackage{upgreek}
\usepackage{hyperref}

\begin{document}

\title{A low maintenance Sr optical lattice clock}

\author{I~R~Hill$^1$, R~Hobson$^{1,2}$, W~Bowden$^{1,2}$, E~M~Bridge$^{1,2}$, S~Donnellan$^1$, E~A~Curtis$^1$ and P~Gill$^1$}

\address{$^1$National Physical Laboratory, Hampton Road, Teddington, TW11~0LW, UK}
\address{$^2$Clarendon Laboratory, University of Oxford, Parks Road, Oxford, OX1 3PU, UK}

\ead{ian.hill@npl.co.uk}

\begin{abstract}
We describe the Sr optical lattice clock apparatus at NPL with particular emphasis on techniques used to increase reliability and minimise the human requirement in its operation.  Central to this is a clock-referenced transfer cavity scheme for the stabilisation of cooling and trapping lasers. We highlight several measures to increase the reliability of the clock with a view towards the realisation of an optical time-scale. The clock contributed 502 hours of data over a 25 day period (84\,\% uptime) in a recent measurement campaign with several uninterrupted periods of more than 48 hours. An instability of $2\times10^{-17}$ was reached after $10^5$\,s of averaging in an interleaved self-comparison of the clock.

\end{abstract}

\section{Introduction}

Ultra-precise optical clocks based on lattice-trapped cold atoms \cite{Baillard2008, Katori2002, Campbell2008, Falke2011, Hong2009} offer a new vision of the SI second with accuracy and precision two orders of magnitude beyond the current realisation \cite{Nicholson2015}. Such precision has profound implications for an Earth-based optical clock network which becomes sensitive to fluctuations in gravitational potential at the 1~cm level. Optical lattice clocks have potential for new applications in numerous precision measurement experiments to determine local gravity \cite{Poli2011}, perform relativistic geodesy and international leveling, form the basis of extremely narrow linewidth lasers \cite{Meiser2009}, provide a route towards quantum computing \cite{Daley2008}, and search for possible time-variation of fundamental constants \cite{Kotochigova2009}.

The reliable, low maintenance operation of these systems is crucial to their adoption in new technologies and in establishing a global optical time-scale. In particular, extended periods of operation place stringent requirements on the frequency stabilisation of multiple laser systems. We report here a simple scheme to transfer the intrinsic stability of the Sr clock to the cooling and trapping laser systems via a single tunable optical cavity; removing the need for regular frequency measurements of the optical lattice trap by femtosecond optical frequency combs, and ensuring a low frequency deviation of the cooling light without additional stabilisation to an atomic vapor. Methods to ensure the precision and robustness of the stabilisation are described, including additional servos for increased laser lock longevity, backed up by automatic relocking algorithms. We also give an overview of the experimental setup at NPL and its first operation as an $^{87}$Sr optical lattice clock.


\section{System overview}\label{system}

\subsection{The cold atom apparatus}\label{chamber}

We adopt a similar approach to others in preparing cold samples of Sr \cite{Katori1999, Loftus2004, Sorrentino2006, Falke2011} using a 2-stage magneto-optical-trap (MOT) loaded from a Zeeman slowed atomic beam.  Details of our oven source and novel transverse-field Zeeman slower, based on a 2D array of permanent magnets, can be found in ref. \cite{Hill2014}.

\begin{figure}[t]
\centering
\includegraphics[width=6.0in]{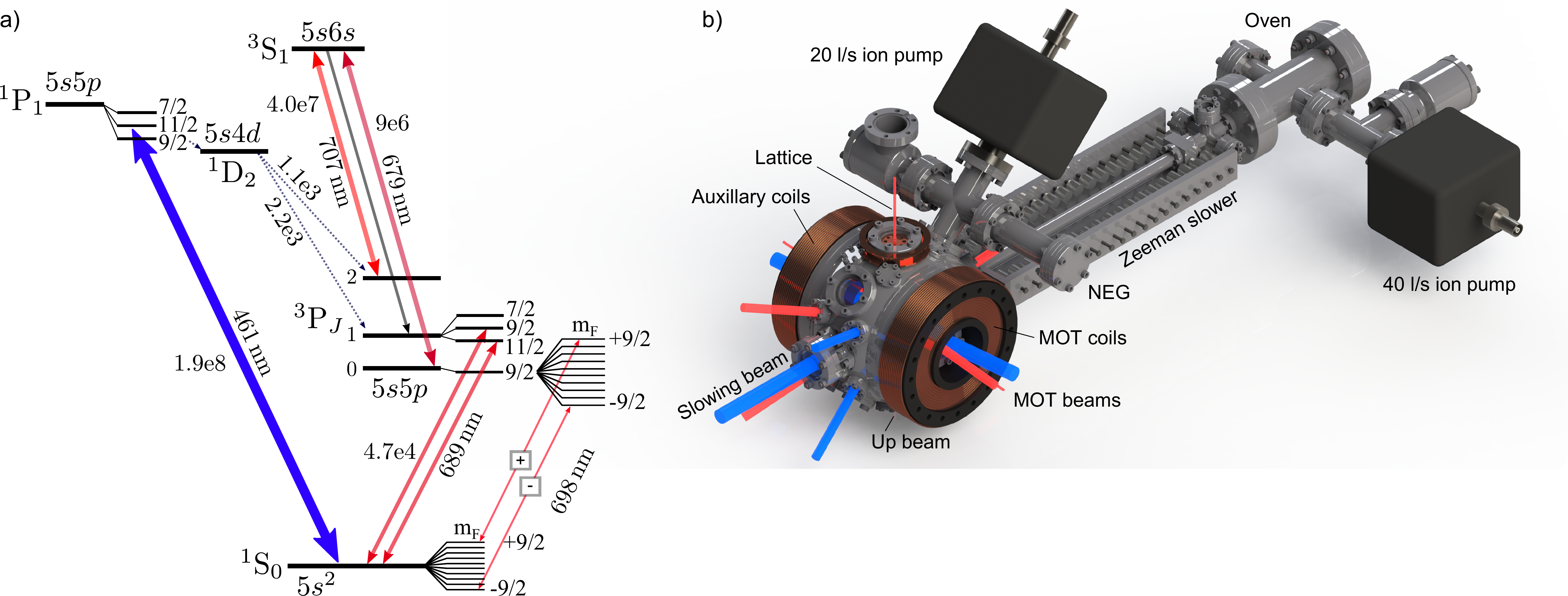}
\caption{(a) The relevant energy levels of Sr for cooling, trapping, and probing the clock transition and (b) a rendering of the cold atom apparatus.
}
\label{fig_chamber}
\end{figure}

The first cooling stage uses the $\gamma=30.2$~MHz broad $^1$S$_0\leftrightarrow\,^1$P$_1$ cycling transition at 461~nm for efficient capture and cooling.  For $^{87}$Sr ($^{88}$Sr) a `blue' MOT formed of 3 retro-reflected beams is loaded for 150--350~ms (60~ms) at saturation intensity of $s= 0.1$ radially and $s= 0.05$ axially with a quadrupole magnetic field gradient of 4~mT\,cm$^{-1}$ and detuning of $-40$~MHz.  The Zeeman slowing beam is then switched off, the MOT detuning is stepped to $-35$~MHz and the intensity ramped down to zero over 15 ms.  This reduces the atomic temperature from around 5~mK to 2~mK and restores the MOT in the quadrupole field zero in preparation for transfer to the second stage MOT.  A branching of 1 in 50,000 from the 5s5p~$^1$P$_1$ to 5s4d~$^1$D$_2$ state necessitates repumping lasers at 707 nm and 679 nm to clear out states 5s5p~$^3$P$_2$ and 5s5p~$^3$P$_0$ via 5s6s~$^3$S$_1$ and return atoms to the cooling cycle via the 5s5p~$^3$P$_1$ state (figure \ref{fig_chamber}a).  

The second cooling stage operates a `red' MOT on the $\gamma=7.5$~kHz narrow $^1$S$_0\leftrightarrow\,^3$P$_1$ intercombination line at 689~nm.  The quadrupole field gradient is switched in $\sim$~1~ms to 0.2~mT\,cm$^{-1}$ to accommodate the weak cooling line.  For $^{87}$Sr a stirring laser operating on the $^1$S$_0|F=9/2\rangle\leftrightarrow\,^3$P$_1|F=9/2\rangle$ line is simultaneously overlapped with all MOT beams to redistribute population amongst $m_{F}$ sublevels for efficient trapping.  The red MOT is initially operated in `broad-band' mode by frequency modulating the 689~nm MOT (stirring) light at 40~kHz with peak-to-peak deviation 4~MHz (2~MHz) to address the Doppler broadened atomic absorption spectrum.  This generates approximately 100 (50) comb lines, with the nearest detuned by $-400$~kHz ($<-100$~kHz), and $s\simeq4$  ($s\simeq3$) per line in a 1~cm $1/e^2$ diameter beam.  Around $10^{5}$ atoms are collected and cooled to $10~\upmu$K during this 60--150~ms phase.  The modulation is then turned off, the intensity jumped to $s\simeq45$ ($s\simeq45$), and the detuning stepped closer to resonance to form a narrow-band MOT.  The cloud is cooled and compressed by ramping the intensity to $s\simeq8$ while reducing the detuning towards resonance over a period of 60~ms.  The intensity is stepped to $s=0.8$ for a final stage of cooling in a low intensity narrow-band MOT reaching temperatures of $\sim$~1~$\upmu$K after 75~ms.  In this final stage an additional `up' beam is needed ($s\simeq30$) to counter gravity which is otherwise compromised by the geometry of the trapping beams.


Atoms are next loaded into a vertical one-dimensional optical lattice trap at 813~nm with $1/e^2$ radius of 45~$\upmu$m.  The lattice is formed by a retro-reflected beam with linear polarisation parallel to the quantisation field direction and up to 1~W from an M-Squared SolTiS Ti:Sapphire laser.  The trap is tuned to near the magic frequency at 368554.5~GHz \cite{Katori2001} and has a $1/e$ lifetime of 8~s.  A spin polarising pulse of circularly polarised 689~nm stirring light, frequency modulated to cover a 2~MHz spectrum, is then applied to transfer population into $m_{F}=\pm9/2$ stretched states in preparation for clock interrogation.  During loading typical trap depths are 75\,$E_\mathrm{r}$ ($\simeq$~12~$\upmu$K), where $E_\mathrm{r}$ is the photon recoil associated with the lattice frequency.  To ensure a low atomic temperature a brief evaporation stage is carried out by ramping the lattice depth to $3~\upmu$K over 20~ms where it remains for a duration of 20 ms before ramping back to the final depth for clock probing.  The clock transition is probed along the vertical axis of the optical lattice in the direction of tight confinement within the Lamb-Dicke regime.  The clock laser has $1/e^2$ beam radius of around 200~$\upmu$m and is linearly polarised along the quantisation field axis.  The power of each cooling and trapping beam is actively stabilised at the fibre output following a polarising cube by feedback to the RF drive power of an acousto-optic modulator (AOM).  Polarising cubes are positioned on entry to each fibre to maintain good alignment with the PM axis and lessen the effect of polarisation fluctuations due to thermal effects in the birefringent TeO$_2$ AOM crystal.



The quadrupole MOT fields are provided by a pair of $9\times8$ turn coils of 4~mm square copper Kapton-insulated wire with a 2.5~mm circular bore through which cooling water flows, and produce a field gradient 50~$\upmu$T\,$\mathrm{{cm}^{-1}}\mathrm{{A}^{-1}}$.  Each coil is potted with Stycast into a Delrin former which provides a 2~mm thick insulation layer between the coil and chamber. Auxiliary coils are wound on to the outer of the former and provide a better approximation to a Helmholtz configuration for quantisation fields and mixing fields for magnetically induced spectroscopy \cite{Taichenachev2006} in $^{88}$Sr, and produce 125~$\upmu$T$\mathrm{{A}^{-1}}$ with long-term instability $<$~20~nT.  MOT (auxiliary) coil currents are adjusted by FET (opamp) actuation in an actively stabilised loop using a Hall (flux gate) sensor.  A bank of transient-voltage-suppressing diodes is used to clamp the switching voltage high to increase the rate of energy transfer from the MOT coils.  The temperature inhomogeneity of the vacuum chamber is maintained at 0.3~K and monitored by 11 Pt-100 sensors calibrated to $\pm10$~mK.

\begin{figure}[t]
\centering
\includegraphics[width=6.35in]{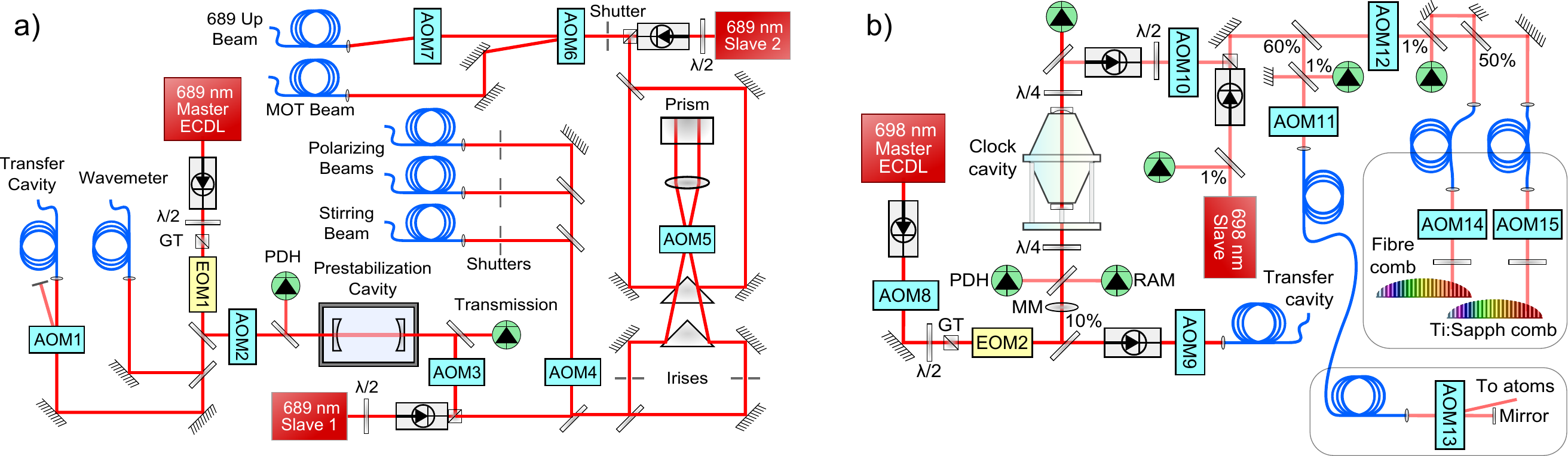}
\caption{(a) 689 nm 2nd-stage cooling laser system and (b) the 698 nm clock laser system. PDH is Pound Drever Hall, GT is Glan Taylor polariser, MM is mode-matching lens, AOM (EOM) is acousto (electro) optic modulator, and ECDL is extended cavity diode laser.}
\label{fig_clock}
\end{figure}

\subsection{Laser apparatus}
\subsubsection{461 nm and repumping.}\label{461}

Around 350~mW of 461~nm light is produced with a commercial frequency-doubled extended cavity diode laser (ECDL) plus tapered amplifier system with bow-tie cavity enhanced second-harmonic generation using a KNbO$_3$ crystal.  The light is split and frequency shifted by AOMs to perform Zeeman slowing, MOT, and absorption and fluorescence imaging probe duties. Repumping is provided by two grating-based ECDLs at 679 nm and 707 nm which are frequency stabilised to a commercial wavemeter via piezo actuation on the cavity length. We modulate the current of the 707 nm diode and individually address all five hyperfine levels of 5s5p~$^3$P$_2$ via separate orders of two AOMs, enabling blue MOT $1/e$ lifetimes $>$ 1~s.

\subsubsection{689 nm cooling, stirring, and spin-polisaring.}\label{689}

The 689~nm laser system shown in figure \ref{fig_clock}a is based on a home-built Littrow-configuration master ECDL seeding two injection-locked slave diode lasers (Opnext HL6750MG). Slave 1 is the source of the light addressing the $F=9/2\rightarrow9/2$ `stirring' and spin-polarising transition, while slave 2 sources the light for the main $f=9/2\rightarrow11/2$ MOT transition. We narrow the linewidth of the master laser by a Pound Drever Hall (PDH) lock to a 60,000-finesse prestabilization cavity (also used to filter the light for injecting the slaves), and we provide long-term frequency stability by a PDH lock to the transfer cavity actuating on the RF drive frequency of AOM2 and 3. The accuracy of the transfer cavity lock is enhanced by implementation of an AOM-based residual-amplitude modulation (RAM) servo (figure~\ref{fig_xfer} and section~\ref{RAMservo}). The frequency of slave 2 can be selected to address either the $^{88}$Sr or $^{87}$Sr MOT transition by opening or closing the irises before AOM5.

\subsubsection{698 nm clock laser.}\label{clocklaser}

The clock laser system shown in figure~\ref{fig_clock}b consists of a home-built Littrow-configuration grating-based ECDL (non-AR-coated Opnext HL7001MG diode) PDH locked to a 7.75~cm long vertical ULE optical cavity \cite{Ludlow2007} whose transmission is used to inject a slave diode laser in a similar scheme to \cite{Sterr2009}. The optical cavity has a measured finesse of 420,000 and bandwidth of 4.6~kHz which provides a roughly 23~dB suppression of the transmitted locking servo sidebands at 1~MHz. The cavity temperature is stabilised ($<1$~mK) close to its minimum of thermal expansion at 25.7~$^{\circ}$C by peltier control of a polished aluminium cylindrical shield isolated within a steel vacuum chamber at 10$^{-7}$ mbar.  Care is taken to avoid residual amplitude modulation by exluding polarising optics in the path between the free-space EOM output and optical cavity input mirror. A Glan-Taylor polarising optic is placed immediately infront of the EOM to ensure a high degree of linear polarisation along the crystal axis. To avoid problematic parasitic etalons and optical feedback to the laser, optical isolation is ensured by insertion of Faraday isolators, multiple beam splitters, and AOM frequency shifting. The complete optical setup is mounted on a two tier $60\times50$~cm active-vibration-isolation platform inside a 12~mm thick acrylic box.  

To extend the locked period of the master laser a low bandwidth servo of the cavity transmission actuating on the laser current is implemented. When the laser is near a mode-hop we observe an increase in cavity transmission due to a change in the modulation character of the diode. Using this change as a discriminator we ensure the mode of the laser is well-tracked.

Light from the clock slave laser is distributed to the atoms and two femtosecond frequency combs labs by path length stabilised fibre links actuated by AOM11/14/15. At the remote end of the atom link AOM13 is used to control the clock light frequency stepping, phase, and amplitude. To eliminate thermally driven phase chirps during switching of the AOM RF drive power, the compensated path is extended to include AOM13. Here, the zero order beam is used in preference to the first order to avoid transient servo capture effects.  


\subsection{Transfer cavity scheme}\label{transfer}

\begin{figure}[h]
\centering
\includegraphics[width=4.6in]{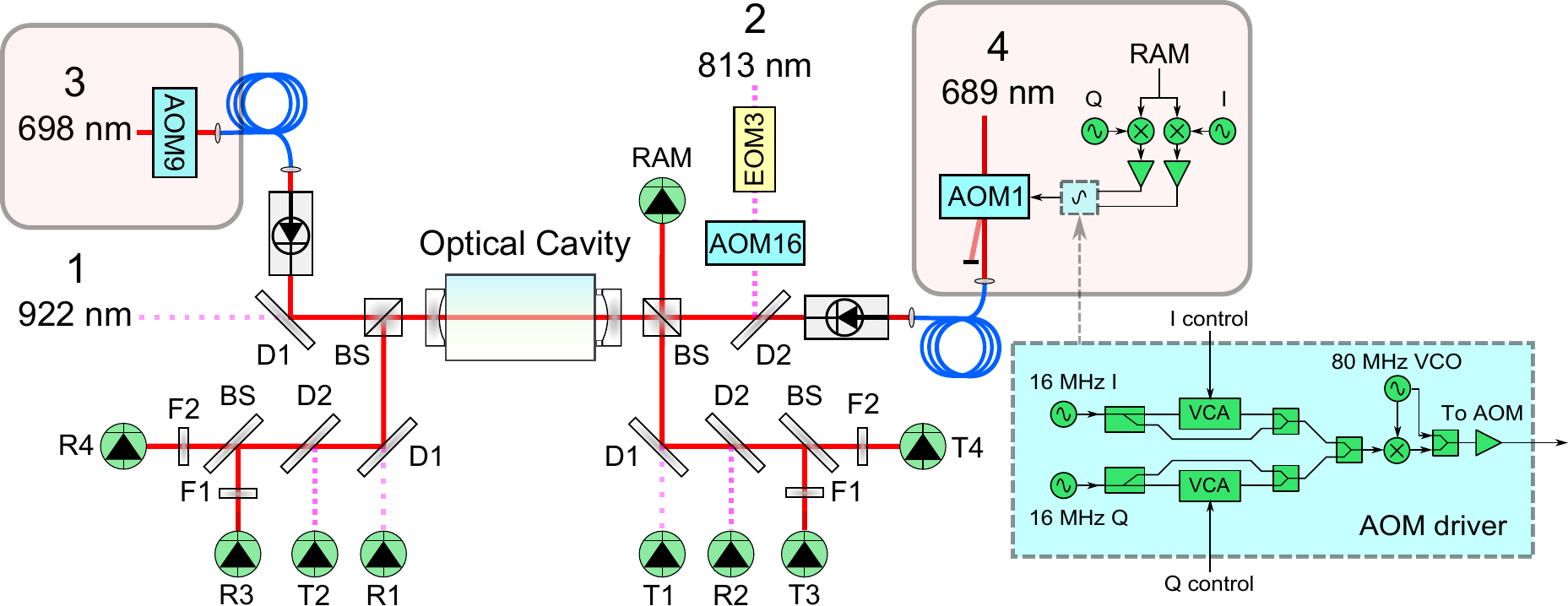}
\caption{The transfer cavity locking scheme. Transmission and reflection photodiodes are labeled T and R and numbered according to wavelength.  BS is a 50:50 beam splitter, D1 is a long pass filter above 900 nm, D2 is a short pass filter below 800 nm, and F1 and F2 are bandpass filters for 698~nm and 689~nm respectively.}
\label{fig_xfer}
\end{figure}

Using a single tunable in-vacuum optical cavity whose length is stabilised to the 698 nm clock laser we transfer the stability of the clock to all cooling and trapping lasers, degraded only by the quality of the individual laser locks to the cavity and the cavity environment.  This removes the need for atomic spectroscopy cells and multiple optical cavities that must be separately characterised.  The scheme, shown in figure~\ref{fig_xfer}, implements a low bandwidth PDH lock at each wavelength with separation by optical filtering (by ensuring sufficiently different PDH modulation frequencies the number of detectors and optical filters could be reduced). AOM9 is updated to track the atomic reference such that the clock cavity drift is removed. The lattice frequency is easily controlled to within $<$~100~kHz as required for 10$^{-18}$ clock uncertainty \cite{Katori2015}.

\subsubsection{AOM residual amplitude modulation servo.}\label{RAMservo}

To allow resolution of the 900~kHz cavity linewidth to the kHz level, the 689~nm light incident on the transfer cavity is picked off onto a separate RAM photodetector and both the in-quadrature (Q) and the in-phase (I) components of RAM are mixed down to DC. Using voltage-controlled attenuator setups that can pass positive- or negative-amplitude RF signals, we can flexibly apply I and Q amplitude modulation to the RF drive carrier of AOM1 (figure \ref{fig_xfer}), diverting a modulated amount of light away from the zero-order beam. We note that the optical fibre after the AOM is an essential component---it spatially filters the beam, homogenising the AM applied by the AOM. Closing the loop, the AM applied to the AOM will exactly compensate the RAM from the EOM (EOM1 figure~\ref{fig_clock}), preventing offsets in the PDH lock. We observe good RAM suppression from typically around $-50$--60~dBm to below $-80$~dBm on our photodetector with a lock bandwidth of a few kHz.  We acknowledge a similar system recently reported in ref~\cite{Hall2015}.

\subsubsection{689 nm frequency correction by lattice loading position servo.}\label{drift}

The stability transfered to the 689~nm laser was at times unsatisfactory, evidenced by deviations in the atom loading position in the optical lattice.  Since the red MOT sags under gravity, its vertical position is proportional to the magnetic field and MOT laser detuning.  Given that the magnetic field is sufficiently stable, we implement a position servo on the centre of mass of the lattice-trapped atom cloud by feeding corrections to AOM9 in preference to those from the clock.  The typical vertical extent of the atomic cloud is about 200~$\upmu$m and is 1:1 imaged by a CCD array with pixel size of $16$~$\upmu$m.  For our loading conditions we obtain a frequency sensitivity of approximately 7~kHz per pixel with sub pixel control resolution.  The residual frequency difference between this method and clock corrections is shown in figure~\ref{fig_clockspec}e.

\subsection{Automated locking algorithms}\label{relocking}

We implement automated injection routines on both 689 nm slave lasers using a commercial wavemeter and knowledge of the laser current injection window. When a slave unlocks, the current is jumped by $\pm1/2$ the injection window. If the slave injects on either of these two test currents, detected by an appropriate reading on the wavemeter, the new operating current will be reset to this value, else the injection window is increased and the process is repeated. Once injected, a similar process is used periodically to ensure the operating current is well centred in the injection window.  

The wavemeter is also used to stabilise and reacquire the lock of both repump lasers.  In the event of a laser mode-hop that is not recovered by the action of the PI servo feeding back to the laser piezo, a lock acquisition routine is entered.  By ramping the piezo and monitoring the laser frequency, an appropriate step is applied to the laser current to move the mode-hop-free window of the ECDL towards the required wavelength where the piezo lock is reinstated.

\subsection{Experimental control}\label{control}
An experimental control system based on bespoke, scalable FPGA hardware has been developed. Scripts which describe the experimental sequence are written in Python and compiled using the Labscript software suite \cite{Starkey2013}, which has been adapted for use with our hardware.  This includes a stackable 8-channel, 16-bit DAC board with a 2~$\upmu$s update period and 24 times 40~MHz digital outputs, accessed via a breakout header board with galvanic isolation. Each DAC board contains an FPGA and 1 Mbit of memory per channel for storing data that is to be clocked out during the experimental shot. A pseudo-clock architecture allows efficient data compression allowing more complex shots to be written and executed. Data acquisition and clock servo operation is by a separate computer based system.

\begin{figure}[t]
\centering
\includegraphics[width=5.5in]{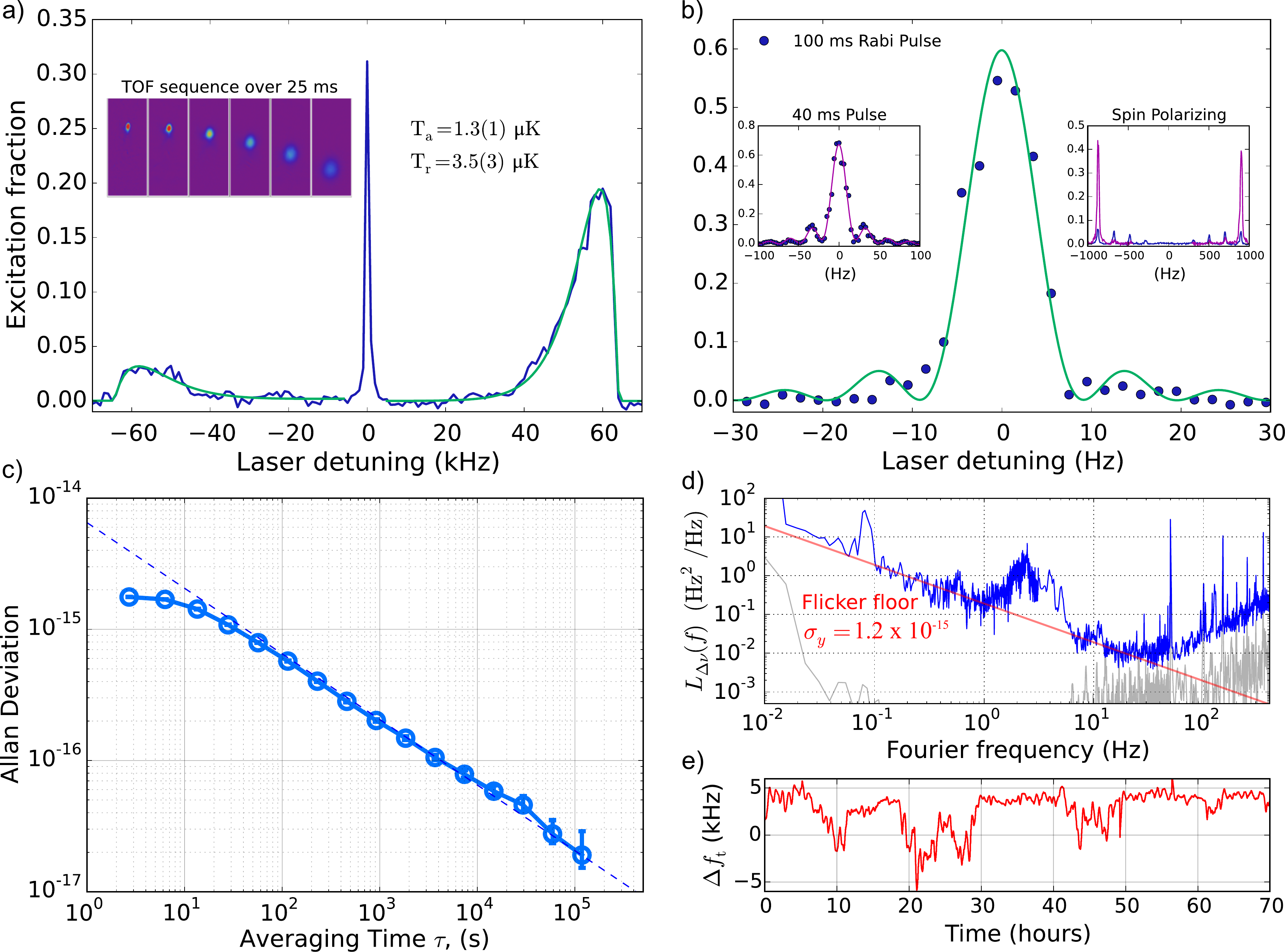}
\caption{(a) Overdriven clock spectroscopy showing first order motional sidebands of lattice-trapped cold Sr and (inset) time-of-flight sequence for 1~$\upmu$K temperature atom cloud released from the red MOT. (b) 100~ms (main) and 40~ms (inset left) Rabi excitation of the spin-polarised clock transition and (inset right) spectroscopy of $m_{F}$ states with and without spin-polarisation. (c) Fractional frequency instability of an interleaved clock self-comparison. (d) Power spectral density of the clock laser frequency noise measured by comparison to an ultra-stable laser at 1064 nm via an optical frequency comb. The bump at 80~mHz is from a temperature oscillation present on the 1064 nm laser system. (e) Frequency corrections made by the lattice loading position servo compared to the clock referenced transfer cavity stabilisation.}
\label{fig_clockspec}
\end{figure}

\section{Performance}\label{spec}
\subsection{Clock spectroscopy}\label{latticespec}


We perform high resolution Rabi spectroscopy of the clock transition in the Lamb-Dicke regime ($\eta \simeq 0.24$) with motion resolved from the carrier.  Spectroscopy of the first-order motional sidebands in an unpolarised sample, shown in figure~\ref{fig_clockspec}a, is used to determine the apparent axial (T$_{\mathrm{a}}$) and radial (T$_{\mathrm{r}}$) temperatures \cite{Blatt2009} and the axial trap frequency for characterisation of lattice shifts. Fourier limited line shapes are achieved for Rabi pulses approaching 100~ms in spin-polarised samples, limited by acoustic noise coupled to the clock laser reference cavity in the region of 2~Hz (see figure~\ref{fig_clockspec}d).

\subsection{Stability and reliability}


To operate the clock we run two servos locked to transitions between the ground and excited $m_{F}=\pm9/2$ stretched states and compute the average frequency $f^{87}=(f^-+f^+)/2$.  The stretched states are split by $\simeq1$~kHz in a 100~$\upmu$T quantisation field.  State populations are read out by normalised shelving detection using a single fluorescence imaging probe at 461~nm.
We achieve an instability of $3.2\times10^{-15}/\sqrt{\tau}$, where $\tau$ is the averaging time, for the non-interleaved clock, measured by an interleaved self-comparison of two $f^{87}$ clock servos with 80~ms Rabi time (see figure \ref{fig_clockspec}c).  An instability of $2\times10^{-17}$ was reached after $10^{5}$\,s of continuous averaging in interleaved operation.  

In a recent 25 day measurement campaign the system recorded 502 hours of `good' frequency data, corresponding to an uptime of 84\,\%. The relocking algorithms described in section~\ref{relocking} were implemented for the last 16 days of the campaign and aided an achieved uptime of 94\,\% for this period. The transfer fidelity of the 698~nm to 689~nm transfer cavity scheme was assessed over a 70 hour period  with reference to the 689~nm frequency servo described in section \ref{drift} (figure~\ref{fig_clockspec}e). We observe a peak deviation of 10~kHz, corresponding to a PDH lock deviation of approximately 1\,\%. An increased low frequency servo gain or increased cavity finesse should yield an improved performance with sub 1~kHz stability transfer.




\section{Conclusion}

We have described a reliable apparatus for a Sr optical lattice clock centred around a transfer cavity laser stabilisation scheme that enables long periods of operation with low maintenance and high accuracy.  The additional servo loops and automatic recapture routines described in this paper enable continuous, reliable clock operation with little human input. Such a system is well placed to contribute to the realisation of an optical time-scale.


\ack
The authors thank Yuri Ovchinnikov, Marco Menchetti, Ross Williams, and Matt Earnshaw for contributions to the setup.  This work was funded by the National Measurement System Electromagnetics and Time programme.

\section*{References}
\bibliography{SrApp}

\end{document}